\documentclass{optica-article}
\journal{opticajournal} 
\articletype{Research Article}

\usepackage{amsmath,amssymb}
\usepackage{bm}
\usepackage{dcolumn}
\usepackage{epstopdf}
\usepackage{lineno}
\usepackage{xcolor}

\begin{document}

\title{Non-Markovian two-time correlation functions for optomechanical systems}

\author{Yifan Shi,\authormark{1} Yusui Chen,\authormark{2,*} and Kaiqi Xiong\authormark{3}}

\address{\authormark{1}Department of Physics, Stevens Institute of Technology, Hoboken, NJ 07030, USA\\
\authormark{2}Physics Department, New York Institute of Technology, Old Westbury, NY 11568, USA\\
\authormark{3}ICNS Lab and Cyber Florida, University of South Florida, Tampa, FL 33620, USA}

\email{\authormark{*}yusui.chen@nyit.edu}

\begin{abstract*}
We extend a stochastic Schr\"odinger equation framework for calculating non-Markovian two-time correlation functions and spectral responses to cavity optomechanical systems. Conventional spectral analysis relies on the quantum regression theorem, whose standard same-generator form is not generally justified for finite-memory environments. In this paper, we consider a linearized optomechanical system in which the mechanical mode interacts with a structured phononic bath and the optical cavity experiences Markovian dissipation. By constructing two-time correlations directly from paired non-Markovian stochastic trajectories, we obtain finite-time spectral functions without invoking regression assumptions. Numerical simulations reveal that environmental memory generates long-lived correlations and qualitatively distinct finite-window spectral features, including non-Lorentzian line shapes and multiple sideband structures relative to the short-memory regime. This work extends the non-Markovian TTCF method to a continuous-variable mixed-bath optomechanical system and provides a framework for studying dynamical correlations in structured environments.
\end{abstract*}

\medskip
\noindent\textbf{Keywords:} open quantum systems; non-Markovian dynamics; stochastic Schr\"{o}dinger equation; two-time correlation function; optomechanics; power spectral density.

\bigskip
\section{Introduction}
\label{sec:intro}

Cavity optomechanics has emerged as a versatile platform for precision measurement and quantum
sensing, enabling ultrasensitive detection of forces, displacements, and external
fields~\cite{OM:aspe2014,OM:NImm2014,OM:Weis1520,OM:otter2018,OM:Brooks2012,OMVitali2007,
OM:Santos:2017,OM:cripe:2018,OM:Arvidsson2024,OM:Wang2024,OM:Hendriks2024,OM:Muffato2025}.
Using the coupling between optical fields and mechanical motion, optomechanical systems support
applications ranging from gravitational-wave detection to quantum-enhanced
metrology~\cite{YChenJPB,Ma2017}.
In these metrological contexts, the central quantity characterizing noise and response is the
two-time correlation function (TTCF), conventionally analyzed using the Wiener--Khinchin theorem,
which relates the power spectral density (PSD) to steady-state autocorrelation
functions~\cite{WK_PRL_2015,WK_2020,WK_2021,WK_QO_2021,WK_2022,WK_2025}:
\begin{equation}
    \mathcal{S}_{AB}(\omega) = \int_{-\infty}^{\infty} d\tau\,
    \langle A(\tau)B(0)\rangle\, e^{-i\omega\tau}.
\end{equation}

\textcolor{black}{This conventional framework is closely connected to the quantum regression theorem (QRT),
which evaluates TTCFs using the same dynamical generator that governs single-time observables~\cite{Lax1963,FordOConnell1996,Guarnieri2014,deVega2017}.
Similarly, the standard input-output formalism widely used in cavity optomechanics is derived
within the Markovian reservoir approximation, where the output field can be related to the
intracavity dynamics through a quantum Langevin description. These approaches have been highly
successful for systems coupled to broadband memoryless environments. However, when the
environment possesses a structured spectrum and a finite correlation time, the system-environment
correlations generated during the evolution cannot in general be captured by a simple
Markovian description. In this regime, non-Markovian memory feeds back into the system dynamics
and modifies both quantum noise and measurement backaction~\cite{Gardiner_1985,Guarnieri2014,deVega2017}.
The presence of additional quantum correlations, such as those introduced by squeezed
environments, further highlights the need for a framework capable of treating environmental
memory and time-dependent TTCFs on equal footing~\cite{squeez-NM}.}

\textcolor{black}{We study a linearized cavity optomechanical system in which the
mechanical mode is coupled to a structured non-Markovian phononic bath, while
the optical cavity undergoes Markovian decay. The analysis is carried out in
the weak-coupling regime and focuses on how finite environmental memory
modifies the two-time correlation functions and the associated finite-time
spectral response.}
Here we develop a stochastic Schr\"{o}dinger equation (SSE) framework\cite{QSD_Gisin84,QSD_Gisin92,QSD_Strunz96,QSD_DiosiStrunz97,
diosi1998,QSD_Strunz99,QSD_Yu99} for 
calculating non-Markovian TTCFs in this 
mixed-bath optomechanical system. By treating the environmental memory through 
time-nonlocal stochastic evolution, the SSE approach provides a two-trajectory 
method for evaluating TTCFs and finite-time spectral functions without relying 
on the quantum regression theorem. This framework enables us to characterize 
how structured phononic environments modify temporal correlations and spectral 
responses beyond the Markovian limit.

The paper is organized as follows.
Section~\ref{sec:model} presents the model Hamiltonian and discusses the bath assumptions.
Section~\ref{sec:method} derives the non-Markovian SSE formalism and the two-trajectory
TTCF method.
Section~\ref{sec:numerical} presents numerical results for different initial-state
preparations and discusses the spectral signatures of non-Markovian dynamics.
Section~\ref{sec:conclusion} provides a summary and outlook.

\section{Model}
\label{sec:model}

We consider a cavity optomechanical system (Figure 1) in which both the mechanical oscillator and the optical
cavity are coupled to their respective environments~\cite{quantum_noise_2002,LOM_1994,LOM_2006,quantum_noise_2010,ShenPRA101_2020,ShenPRA98_2018,ShenPRA109_2024,
ShenPRA97_2018,ShenPRA96_2017,LiuPRA87_2013}.
The total Hamiltonian is $H_{\rm tot} = H_{\rm sys} + H_{\rm int} + H_{\rm env}$.
In the interaction picture with respect to the environments and with $\hbar = 1$:
\begin{eqnarray}
    H_{\rm sys} &=& \omega_0 a^\dagger a + \Omega b^\dagger b
        - \lambda(a^\dagger + a)(b^\dagger + b),
    \nonumber\\
    H_{\rm int} &=& b\sum_k g_k^* e_k^\dagger e^{i\omega_k t}
                  + a\sum_j h_j^* c_j^\dagger e^{i\nu_j t} + \mathrm{H.c.},
    \label{eq:Htot}
\end{eqnarray}
where $a$ ($a^\dagger$) and $b$ ($b^\dagger$) are the cavity and mechanical bosonic operators,
$\omega_0$ and $\Omega$ are the corresponding mode frequencies, and $\lambda$ is the effective
linearized optomechanical coupling~\cite{LOM_1994,LOM_2006}. The operators $e_k$ ($e_k^\dagger$)
and $c_j$ ($c_j^\dagger$) describe phononic and photonic bath modes with frequencies $\omega_k$
and $\nu_j$, respectively, and coupling strengths $g_k$ and $h_j$. The system--bath coupling
operators are $L_1=b$ for the phononic bath and $L_2=a$ for the photonic bath.

\begin{figure}[htbp]
    \centering
    \includegraphics[width=0.42\linewidth]{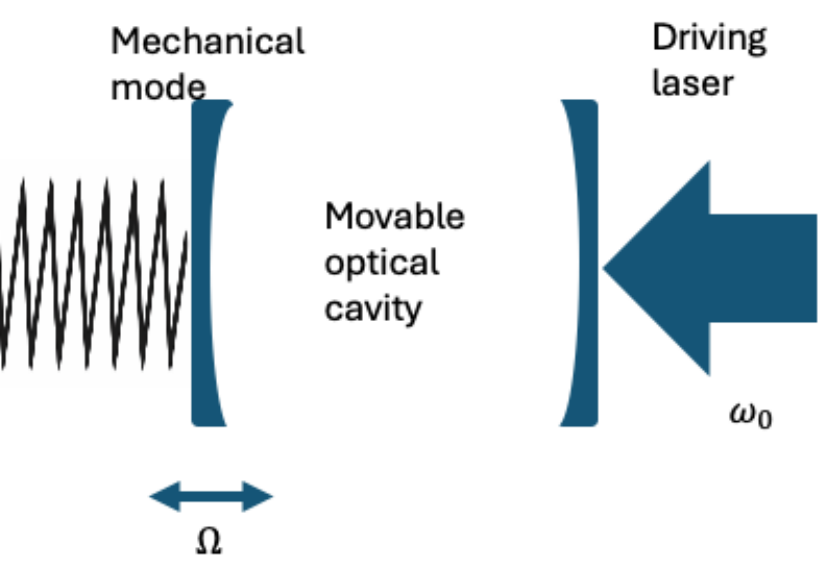}
    \caption{Schematic of the cavity optomechanical system. A mechanical resonator (eigenfrequency
    $\Omega$) couples to the optical cavity (resonance $\omega_0$) via the effective linearized optomechanical coupling ($\lambda$).
    The mechanical mode is coupled to a non-Markovian phononic bath ($L_1 = b$, memory parameter
    $\gamma$, coupling strength $\Gamma$); the optical mode is coupled to a Markovian photonic bath
    ($L_2 = a$, decay rate $\kappa$).}
    \label{fig:sketch}
\end{figure}

The two baths are characterized by different memory properties. The phononic bath has a
non-Markovian Ornstein--Uhlenbeck (OU) correlation function:
\begin{equation}
    \alpha_{\rm ph}(t,s) = \frac{\Gamma\gamma}{2}\,e^{-\gamma|t-s|},
    \label{eq:OU}
\end{equation}
where $\gamma$ is the inverse memory time ($\gamma \to \infty$ is the Markov limit) and $\Gamma$
is the coupling strength. The photonic bath is taken in the Markovian (white-noise) limit:
\begin{equation}
    \alpha_{\rm opt}(t,s) = \kappa\,\delta(t-s),
    \label{eq:white}
\end{equation}
where $\kappa$ is the cavity decay rate~\cite{OM_NM_2017,NM:OM:Direkci2024,squeez-NM}.
The Markovian approximation for the optical bath is well established.
Cavity photon loss is mediated by mirror transmission into a broadband electromagnetic
continuum.  The correlation time of this photonic environment is set by the inverse bandwidth
of the mirror reflectivity, typically of order femtoseconds---orders of magnitude shorter than
the cavity lifetime $\kappa^{-1}$ (microseconds to milliseconds in typical optomechanical
experiments) and the mechanical period $\Omega^{-1}$.
This is precisely the regime in which the Born--Markov approximation for the photonic bath is
exact, and it underpins the standard input--output
theory~\cite{Gardiner_1985,quantum_noise_2002,quantum_noise_2010} that describes cavity
photon loss so successfully. The resulting Markovian master equation for the cavity field,
with Lindblad dissipator $\kappa\mathcal{D}[a]\rho$, is validated by cavity-QED
and optomechanical experiments~\cite{OM:aspe2014}.

The mechanical bath, by contrast, can be genuinely non-Markovian.
Mechanical oscillators in cryogenic optomechanical experiments dissipate energy into
a phononic environment whose spectral density is not flat: crystalline substrates have
Debye-type acoustic spectra with structure on the scale of the mechanical
frequency~\cite{quantum_noise_2002}, and amorphous materials harbour two-level system
defects that produce strongly colored (non-Lorentzian) noise at low
temperatures~\cite{quantum_noise_2010}.
For high-$Q$ mechanical resonators, the bath correlation time $\gamma^{-1}$ can be comparable
to $\Omega^{-1}$, making memory effects observable in the spectral response and
TTCFs~\cite{OM_NM_2017,NM:OM:Direkci2024}.
The OU kernel~(\ref{eq:OU}) captures this non-Markovian structure through the single parameter
$\gamma$: small $\gamma$ corresponds to long bath memory (strongly non-Markovian), while
$\gamma \to \infty$ recovers the Markovian limit.


\section{Stochastic Schr\"{o}dinger-equation method for TTCFs}
\label{sec:method}

\subsection{Quantum state diffusion formalism}

Since the optical environment has already been taken in the Markovian limit, its contribution to the reduced system dynamics is
described by
\begin{equation}
H_{\rm eff}
=
H_{\rm sys}
-\frac{i\kappa}{2}a^\dagger a.
\label{eq:Heff}
\end{equation}
The non-Hermitian Hamiltonian $H_{\rm eff}$ governs the deterministic
no-jump branch of the Markovian optical evolution. In the trajectory construction
used below, the optical decay is included through the conditional propagator
generated by $H_{\rm eff}$, whereas the phononic environment is retained
explicitly through the non-Markovian quantum-state diffusion (QSD) representation.

The conditional state of the cavity--mechanical system and the phononic
environment then satisfies
\begin{equation}
\partial_t|\Psi(t)\rangle
=
-i\left[
H_{\rm eff}
+
H_{{\rm int},{\rm ph}}(t)
\right]
|\Psi(t)\rangle,
\label{eq:conditional_SE}
\end{equation}
where
\begin{equation}
H_{{\rm int},{\rm ph}}(t)
=
b\sum_k g_k^*e_k^\dagger e^{i\omega_k t}
+
b^\dagger\sum_k g_k e_k e^{-i\omega_k t}.
\label{eq:H_int_ph}
\end{equation}
The corresponding conditional propagator is denoted by $\tilde U(t)$. The
initial state is taken to be
\begin{equation}
|\Psi(0)\rangle
=
|\psi_s(0)\rangle\otimes|0\rangle_{\rm ph},
\label{eq:initial_state}
\end{equation}
where $|\psi_s(0)\rangle$ is the initial cavity--mechanical state. It may be
either separable or correlated with respect to the cavity and mechanical
degrees of freedom.

To derive the non-Markovian QSD equation, we expand the phononic environment
in the unnormalized Bargmann coherent-state basis
\begin{equation}
|z\rangle_{\rm ph}
=
\bigotimes_k
e^{z_ke_k^\dagger}|0_k\rangle.
\label{eq:bargmann_state}
\end{equation}
These states satisfy
\begin{equation}
{}_{\rm ph}\langle z|e_k^\dagger
=
z_k^*\,{}_{\rm ph}\langle z|,
\qquad
{}_{\rm ph}\langle z|e_k
=
\frac{\partial}{\partial z_k^*}
{}_{\rm ph}\langle z|,
\label{eq:bargmann_action}
\end{equation}
and the completeness relation
\begin{equation}
\int d\mu(z)\,
|z\rangle_{\rm ph}\langle z|_{\rm ph}
=
\mathbb I_{\rm ph},
\label{eq:bargmann_identity}
\end{equation}
where
\begin{equation}
d\mu(z)
=
\prod_k\frac{d^2z_k}{\pi}e^{-|z_k|^2}.
\label{eq:bargmann_measure}
\end{equation}

The stochastic cavity--mechanical state is defined by projecting the
conditional state onto the phononic coherent-state basis,
\begin{equation}
|\psi_z(t)\rangle
=
{}_{\rm ph}\langle z|\Psi(t)\rangle
=
{}_{\rm ph}\langle z|
\tilde U(t)
|\Psi(0)\rangle.
\label{eq:psi}
\end{equation}
Applying Eq.~(\ref{eq:bargmann_action}) to
Eq.~(\ref{eq:conditional_SE}) gives
\begin{equation}
\partial_t|\psi_z(t)\rangle
=
\left[
-iH_{\rm eff}
+
bz_t^*
-
i b^\dagger
\sum_k g_k e^{-i\omega_k t}
\frac{\partial}{\partial z_k^*}
\right]
|\psi_z(t)\rangle,
\label{eq:projected_SE}
\end{equation}
where the stochastic noise is
\begin{equation}
z_t^*
=
-i\sum_k g_k^*z_k^*e^{i\omega_k t}.
\label{eq:noise}
\end{equation}
Under the Gaussian measure in Eq.~(\ref{eq:bargmann_measure}), the Bargmann
variables satisfy
\begin{equation}
\mathcal M[z_kz_l^*]
=
\delta_{kl},
\end{equation}
and therefore
\begin{equation}
\mathcal M[z_tz_s^*]
=
\sum_k|g_k|^2e^{-i\omega_k(t-s)}
=
\alpha_{\rm ph}(t,s),
\label{eq:noise_corr}
\end{equation}
where $\mathcal M[\cdot]$ denotes the ensemble average.

The discrete index $k$ labels the individual phononic bath modes, whereas
$s$ denotes the continuous time variable of the resulting noise history.
These two descriptions are related through the functional chain rule
\begin{equation}
\frac{\partial}{\partial z_k^*}
=
\int_0^t ds\,
\frac{\partial z_s^*}{\partial z_k^*}
\frac{\delta}{\delta z_s^*},
\label{eq:noise_chain_rule}
\end{equation}
with
\begin{equation}
\frac{\partial z_s^*}{\partial z_k^*}
=
-i g_k^*e^{i\omega_k s}.
\end{equation}
It follows that
\begin{equation}
-i\sum_k g_k e^{-i\omega_k t}
\frac{\partial}{\partial z_k^*}
=
-\int_0^t ds\,
\alpha_{\rm ph}(t,s)
\frac{\delta}{\delta z_s^*}.
\end{equation}
Equation~(\ref{eq:projected_SE}) therefore becomes
\begin{equation}
\partial_t|\psi_z(t)\rangle
=
\left[
-iH_{\rm eff}
+
bz_t^*
-
b^\dagger
\int_0^t ds\,
\alpha_{\rm ph}(t,s)
\frac{\delta}{\delta z_s^*}
\right]
|\psi_z(t)\rangle.
\label{eq:qsd_functional}
\end{equation}

The functional derivative is represented by the $O$ operator,
\begin{equation}
\frac{\delta}{\delta z_s^*}
|\psi_z(t)\rangle
=
O(t,s)|\psi_z(t)\rangle,
\label{eq:O_operator}
\end{equation}
and its bath-weighted form
\begin{equation}
\bar O(t)
=
\int_0^t ds\,
\alpha_{\rm ph}(t,s)O(t,s).
\label{eq:O_bar}
\end{equation}
The operator $O(t,s)$ describes the response of the stochastic state at time
$t$ to a variation of the noise at the earlier time $s$, while $\bar O(t)$
contains the accumulated effect of the phononic memory. The QSD equation can
then be written in the time-local form
\begin{equation}
\partial_t|\psi_z(t)\rangle
=
\left[
-iH_{\rm eff}
+
bz_t^*
-
b^\dagger\bar O(t)
\right]
|\psi_z(t)\rangle.
\label{eq:qsd_general}
\end{equation}
This construction follows directly from the phononic Bargmann projection and
retains the continuous environmental-noise history
~\cite{QSD_Gisin84,QSD_Gisin92,QSD_Strunz96,QSD_Strunz99,QSD_Yu99}.

Within this trajectory description, the two-time correlation function
considered in this work is
\begin{equation}
\langle A(t)B\rangle
=
{\rm Tr}_{S\otimes{\rm ph}}
\left[
A\tilde U(t)B
|\Psi(0)\rangle\langle\Psi(0)|
\tilde U^\dagger(t)
\right],
\label{eq:ttcf}
\end{equation}
where $A$ and $B$ are cavity--mechanical system operators. Initial
cavity--mechanical correlations can enter this quantity through nonzero cross
moments and can consequently modify its finite-window spectral transform.

Inserting the phononic Bargmann completeness relation gives
\begin{equation}
\langle A(t)B\rangle
=
{\rm tr}_S
\left[
A
\int d\mu(z)\,
{}_{\rm ph}\langle z|
\tilde U(t)B|\Psi(0)\rangle
\langle\Psi(0)|
\tilde U^\dagger(t)
|z\rangle_{\rm ph}
\right].
\label{eq:ttcf_phi}
\end{equation}
This expression provides the starting point for the paired-trajectory
construction introduced below~\cite{carballeira_stochastic_2021,Shi2023}.

\subsection{The $O$-operator and physical interpretation}

With the definition of the $O$-operator, its evolution equation follows from the consistency condition
$\partial_t(\delta_{z_s^*}|\psi_z\rangle) = \delta_{z_s^*}(\partial_t|\psi_z\rangle)$~\cite{QSD_Strunz99,chen_non-markovian_2018}:
\begin{equation}
    \partial_t O(t,s) = \left[-iH_{\rm eff} - L_1^\dagger\bar{O},\; O\right].
    \label{eq:Oevo}
\end{equation}

For the linearized Hamiltonian~(\ref{eq:Heff}) and $L_1 = b$, we adopt the
ansatz~\cite{chen_non-markovian_2018}:
\begin{equation}
    O(t,s) = f_1(t,s)\,a + f_2(t,s)\,a^\dagger + f_3(t,s)\,b + f_4(t,s)\,b^\dagger,
    \label{eq:Oansatz}
\end{equation}
with boundary conditions $f_3(t,t)=1$ and $f_{1,2,4}(t,t)=0$.
\textcolor{black}{Each coefficient in the $O$-operator expansion represents a distinct
operator channel associated with the coupled cavity--mechanical dynamics. The
coefficient $f_3$ corresponds to the direct mechanical channel coupled to the
phononic environment, as indicated by the boundary condition
$f_3(t,t)=1$. Its deviation from the Markovian limit reflects the memory
effect of the structured bath. }

\subsection{Coefficient-function ODE system}

Substituting ansatz~(\ref{eq:Oansatz}) into Eq.~(\ref{eq:Oevo}), the term
$-iH_{\rm eff} = -iH_s - \frac{\kappa}{2}a^\dagger a$ contributes to $\dot{O}$ via:
\begin{equation}
    \left[-\tfrac{\kappa}{2}a^\dagger a,\; f_1 a\right]
    = -\tfrac{\kappa}{2}f_1[a^\dagger a,\, a] = +\tfrac{\kappa}{2}f_1\, a,
    \qquad
    \left[-\tfrac{\kappa}{2}a^\dagger a,\; f_2 a^\dagger\right]
    = -\tfrac{\kappa}{2}f_2[a^\dagger a,\, a^\dagger] = -\tfrac{\kappa}{2}f_2\, a^\dagger,
\end{equation}
using $[a^\dagger a, a] = -a$ and $[a^\dagger a, a^\dagger] = +a^\dagger$, while
$[a^\dagger a,\, b^{(\dagger)}] = 0$ leaves $f_3$ and $f_4$ unaffected.
Defining $F_j(t) = \int_0^t ds\,\alpha_{\rm ph}(t,s)f_j(t,s)$ and applying the OU identity
$\dot{F}_j = \frac{\Gamma\gamma}{2}f_j(t,t) - \gamma F_j + \int_0^t \alpha_{\rm ph}\,\dot{f}_j\, ds$,
the four coupled ODEs are:
\begin{eqnarray}
    \dot{F}_1 &=& \left(-\gamma + i\omega_0 + \tfrac{\kappa}{2}\right)F_1
                  - i\lambda(F_3 - F_4) + F_1 F_3,
    \label{eq:F1}\\
    \dot{F}_2 &=& \left(-\gamma - i\omega_0 - \tfrac{\kappa}{2}\right)F_2
                  - i\lambda(F_3 - F_4) + F_2 F_3,
    \label{eq:F2}\\
    \dot{F}_3 &=& \tfrac{\Gamma\gamma}{2} + (-\gamma + i\Omega)F_3
                  - i\lambda(F_1 - F_2) + F_3^2,
    \label{eq:F3}\\
    \dot{F}_4 &=& (-\gamma - i\Omega)F_4
                  - i\lambda(F_1 - F_2) + F_3 F_4,
    \label{eq:F4}
\end{eqnarray}
with $F_j(0)=0$. 
\textcolor{black}{The reduction to the time-local equations
Eqs.~(\ref{eq:F1})--(\ref{eq:F4}) makes use of the exponential form of the OU
bath correlation function. The same construction can be extended to a more
general structured environment whose correlation function is represented by a
finite sum of exponential terms~\cite{YuFiniteT2004,RitschelEisfeld2014},
\begin{equation}
\alpha_{\rm ph}(t,s)
\simeq
\sum_{r=1}^{R}
c_r e^{-\gamma_r(t-s)}.
\label{eq:multi_expansion}
\end{equation}
In that case, one introduces a separate set of memory-integrated coefficients
for each exponential component, and the total $\bar O$ operator is obtained by
summing their contributions. The resulting enlarged set of coupled equations
has the same general structure as
Eqs.~(\ref{eq:F1})--(\ref{eq:F4}).}

The $\kappa$-dependent terms in Eqs.~(\ref{eq:F1}) and
(\ref{eq:F2}) should be interpreted at the level of the auxiliary
$O$-operator parametrization. In particular, the linear part of the $F_1$
equation contains the real coefficient
\begin{equation}
-\gamma+\frac{\kappa}{2}.
\end{equation}
When $\kappa>2\gamma$, this isolated linear contribution has a positive real
part. \textcolor{black}{This observation alone, however, does not imply physical amplification
or instability of the complete cavity and mechanical system. The coefficient
$F_1$ is an auxiliary memory function rather than a directly measurable
quantity, and it evolves as part of the nonlinear coupled system
Eqs.~(\ref{eq:F1})--(\ref{eq:F4}). Its asymptotic fixed points and possible
singular structures require a separate nonlinear stability analysis, which is
outside the scope of the present work.}

For the parameter regime considered below, direct numerical integration shows
that the four auxiliary coefficients remain bounded and approach stable
long-time values. To keep the discussion focused on the non-Markovian TTCFs
and their spectral consequences, we use $\kappa=0.1\Omega$ in the numerical calculations unless otherwise specified.

\subsection{Two-trajectory method for TTCF}

Because the Markovian photonic bath has been absorbed into $H_{\rm eff}$, the only sampled noise
is the phononic noise. The TTCF therefore reduces to a single Bargmann
average~\cite{carballeira_stochastic_2021,Shi2023}:
\begin{equation}
    \langle A(t)B\rangle = \mathrm{tr}_S\!\left[
    A\int d\mu(z)\,\langle z|_{\rm ph}\,\tilde{U}\,B|\psi_s(0)\rangle
    \langle\psi_s(0)|\,\tilde{U}^\dagger|z\rangle_{\rm ph}\right].
\end{equation}
The term $\langle z|_{\rm ph}\,\tilde{U}\,B|\psi_s(0)\rangle$ has the same structure as the
stochastic trajectory $|\psi_z\rangle$ but with the initial condition replaced by
$B|\psi_s(0)\rangle$.
We therefore introduce a normalized second trajectory~\cite{Shi2023}:
\begin{equation}
    |\phi_z\rangle = \frac{\langle z|_{\rm ph}\,\tilde{U}\,B|\psi_s(0)\rangle}
                         {\sqrt{\langle B^\dagger B\rangle_0}},
    \qquad
    \langle B^\dagger B\rangle_0 = \mathrm{tr}_S\!\left[
    B|\psi_s(0)\rangle\langle\psi_s(0)|B^\dagger\right],
    \label{eq:phi}
\end{equation}
which obeys Eq.~(\ref{eq:qsd_general}) with initial condition
$|\phi_z(0)\rangle = B|\psi_s(0)\rangle$.
\textcolor{black}{The TTCF is then:
\begin{equation}
    \langle A(t)B\rangle = \sqrt{\langle B^\dagger B\rangle_0} \cdot
    \mathrm{tr}_S\!\left[A\,\mathcal{M}(|\phi_z\rangle\langle\psi_z|)\right].
    \label{eq:ttcf_final}
\end{equation}}
For each Bargmann sample, the two trajectories are propagated using the same
phononic noise history $z_t^*$, because they originate from the same coherent-state
projection $\langle z|_{\rm ph}$ of the phononic environment. The TTCF must therefore
be constructed from correlated trajectory pairs. Using independent noise
realizations for $|\phi_z(t)\rangle$ and $|\psi_z(t)\rangle$ would no longer
represent the common environmental trace appearing in
Eq.~(\ref{eq:ttcf_phi}).

The ensemble average is evaluated as
\begin{equation}
\mathcal{M}\!\left(
|\phi_z(t)\rangle\langle\psi_z(t)|
\right)
=
\frac{1}{N_{\rm traj}}
\sum_{n=1}^{N_{\rm traj}}
|\phi_{z_n}(t)\rangle
\langle\psi_{z_n}(t)| ,
\end{equation}
where each pair is generated from the same noise realization $z_n$. Similarly,
the single-trajectory average
\begin{equation}
\rho_r = \mathcal{M}\!\left(
|\psi_z(t)\rangle\langle\psi_z(t)|
\right)
\end{equation}
gives the reduced system density matrix after averaging over the phononic environment. In the
present treatment, where the photonic bath is incorporated through the
non-Hermitian effective Hamiltonian, this state corresponds to the conditional
optical no-jump evolution. The unconditional reduced density matrix requires
the full Markovian optical channel.

\section{Numerical results}
\label{sec:numerical}

\subsection{Coefficient functions of the \texorpdfstring{$\bar{O}$}{Obar} operators}

\begin{figure}[htbp]
    \centering
    \includegraphics[width=0.90\linewidth]{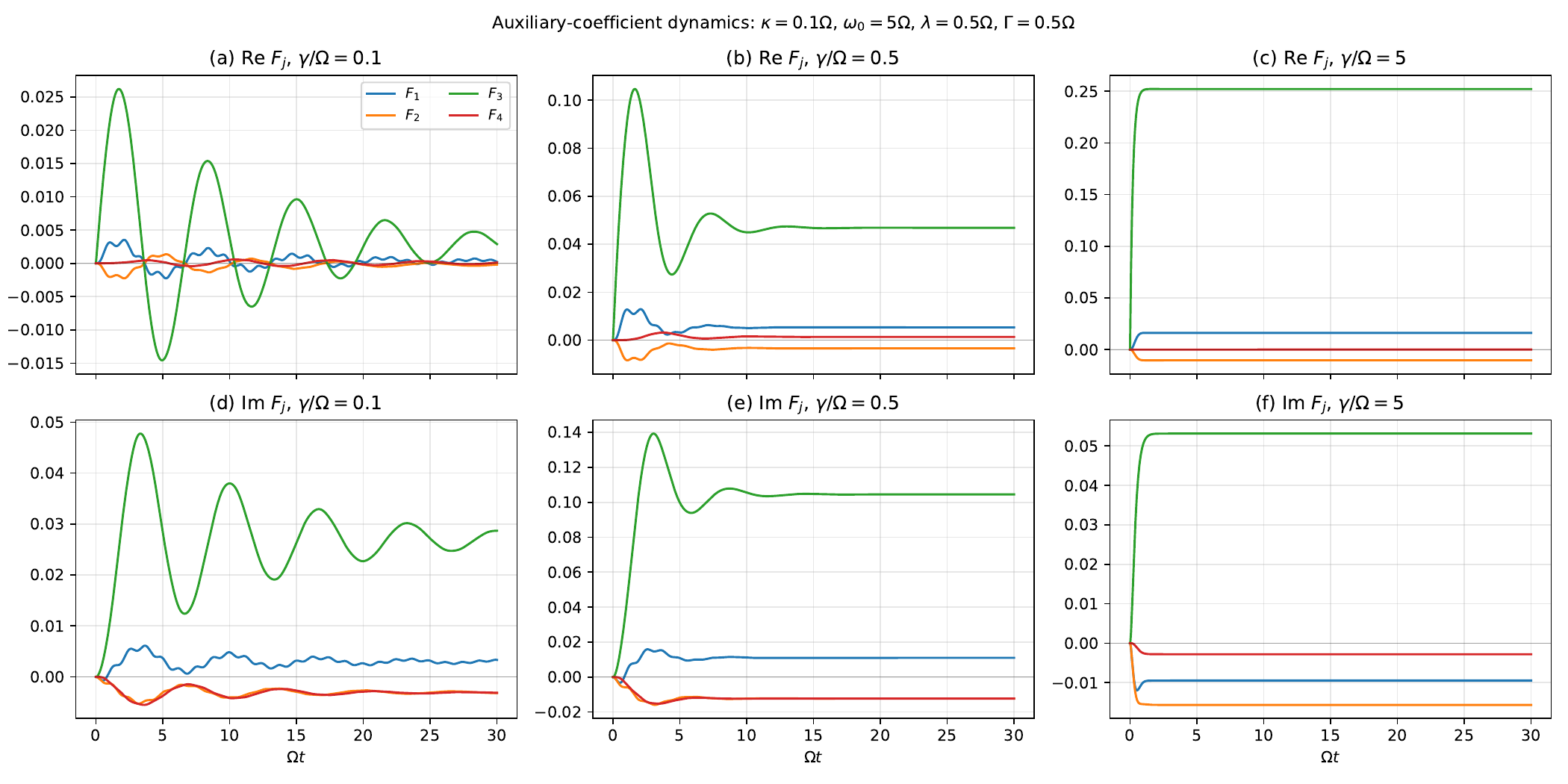}
    \caption{Dynamics of the auxiliary $\bar O$ coefficients. The three columns correspond to $\gamma/\Omega=0.1$, $0.5$, and $5$; the upper and lower rows show the real and imaginary parts, respectively. Other parameters are $\kappa=0.1\Omega$, $\omega_0=5\Omega$, $\lambda=0.5\Omega$, and $\Gamma=0.5\Omega$. The curves describe auxiliary closure coordinates, not separately observable dynamical variables.}
    \label{fig:coeff}
\end{figure}

Figure~\ref{fig:coeff} shows the time evolution of the phononic coefficients $F_1$--$F_4$.
\textcolor{black}{Unless stated otherwise, the calculations below use $\kappa=0.1\Omega$.
Figure~\ref{fig:coeff} compares $\gamma/\Omega=0.1$, $0.5$, and $5$ at $\omega_0=5\Omega$,
$\lambda=0.5\Omega$, and $\Gamma=0.5\Omega$, so that only the bath memory rate is varied.}


In the non-Markovian regime ($\gamma=0.2$), \textcolor{black}{the auxiliary coefficients exhibit
long-lived damped oscillations associated with the finite memory time of the
phononic environment. In the short-memory regime ($\gamma=5$), these oscillations
decay more rapidly and the coefficients approach finite long-time values. The
coefficients $F_j$ enter the memory-integrated operator $\bar O(t)$ and thereby
contribute to the stochastic evolution and the TTCF calculation. They are
auxiliary quantities of the $O$-operator representation rather than separately
measurable observables.}

\subsection{Fock-state cavity, mechanical vacuum}
\label{sec:fock}

The displacement and force-quadrature operators are $x = b + b^\dagger$ and
$\hat{F} = a + a^\dagger$, with $\lambda\hat{F}$ proportional to the linearized
radiation-pressure force on the mechanical mode. Their TTCFs and finite-time PSDs are defined as
\begin{equation}
    C_{xx}(\tau) = \langle x(\tau)x(0)\rangle,\quad
    S_{xx}(\omega) = \int_0^T d\tau\,e^{i\omega\tau}C_{xx}(\tau),
\end{equation}
\begin{equation}
    C_{FF}(\tau) = \langle \hat{F}(\tau)\hat{F}(0)\rangle,\quad
    S_{FF}(\omega) = \int_0^T d\tau\,e^{i\omega\tau}C_{FF}(\tau),
\end{equation}
where $T$ is the finite observation time chosen sufficiently large that Markovian dynamics have relaxed.

The TTCFs considered in this subsection are defined without
subtracting the first moments. For the factorized Fock-state preparation
$|\psi_s(0)\rangle=|1\rangle_c\otimes|0\rangle_m$, the initial quadrature
means vanish,
$\langle x(0)\rangle=\langle\hat{F}(0)\rangle=0$.
Accordingly, $S_{FF}^{(T)}$ is referred to as a force-quadrature finite-window
spectrum rather than a stationary force-noise spectrum.

In a stationary linear-response setting, the force-driven component of
the displacement spectrum may be written as
$S_{xx}^{\rm driven}(\omega)
=|\chi_{xF}(\omega)|^2S_{FF}(\omega)$,
provided that intrinsic mechanical fluctuations and force--displacement cross
correlations are treated separately~\cite{YChenJPB,bbli_2021, xia_2023, zhang_2024}.
The quantity $S_{xx}^{(T)}$ evaluated here is instead a finite-window transform
of the total displacement TTCF, for which no stationary limit is assumed.
This factorization therefore does not generally apply. We calculate
$S_{xx}^{(T)}$ and $S_{FF}^{(T)}$ independently and use their quotient only as
a descriptive spectral ratio, rather than as an estimate of the mechanical
susceptibility. The ratio is evaluated only where its independently calculated denominator has appreciable support:
\begin{equation}
\mathcal R_T(\omega)=\frac{\widehat S_{xx}^{(T)}(\omega)}
{\widehat S_{FF}^{(T)}(\omega).}
\end{equation}


We initialize the cavity in the Fock state $|\psi_c(0)\rangle = |1\rangle$ and the
mechanical mode in vacuum $|\psi_m(0)\rangle = |0\rangle$. The phononic bath is of OU form
with $\Gamma = 0.5\Omega$, and the system parameters are $\omega_0 = 5\Omega$ and $\lambda = 1$.


\begin{figure}[htbp]
\centering
\begin{minipage}[t]{0.48\linewidth}
    \centering
    \captionsetup{width=\linewidth}
    \includegraphics[width=\linewidth]{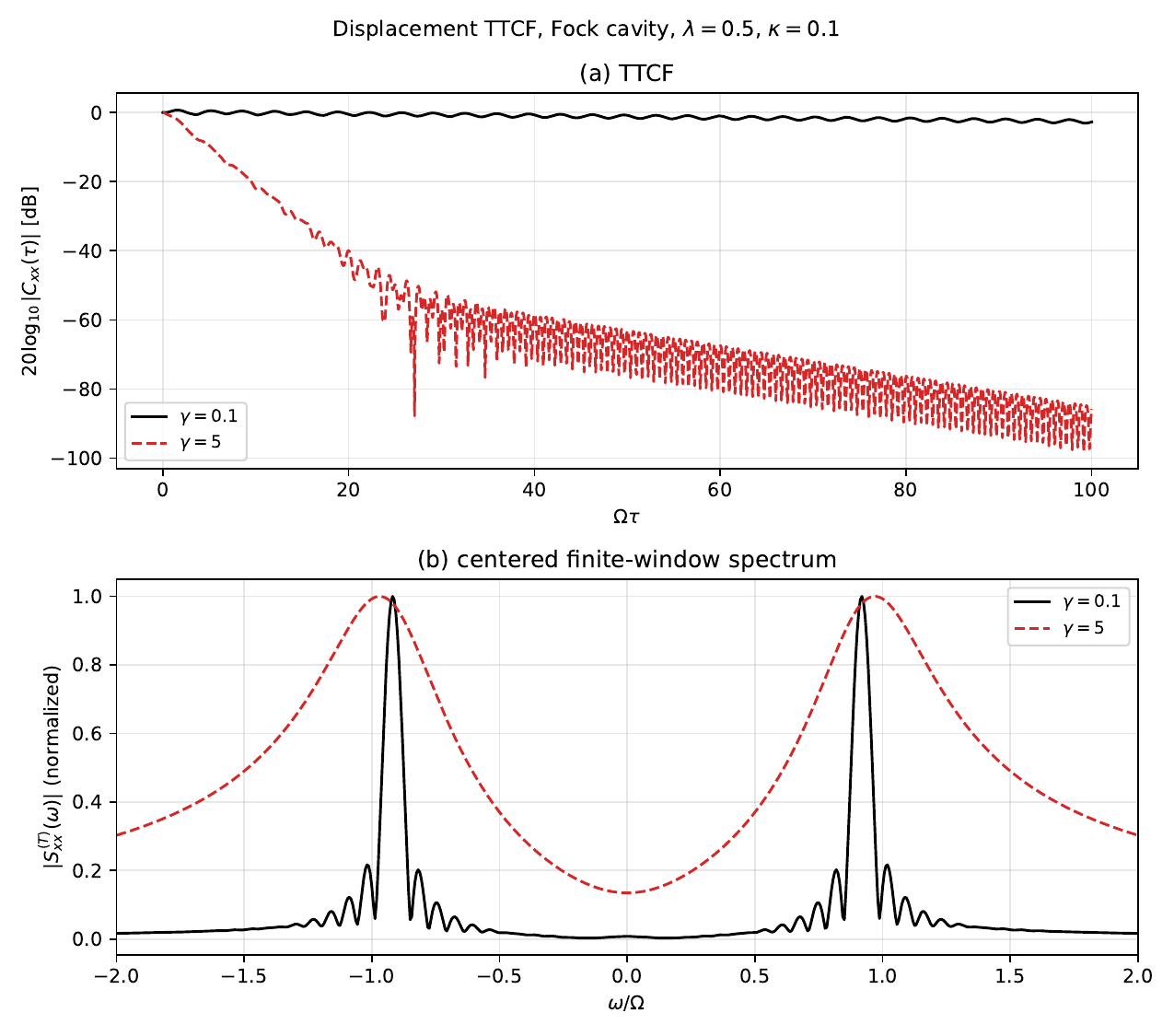}
    \captionof{figure}{%
    Displacement panel: $20\log_{10}|C_{xx}(\tau)|$ (a) and the normalized, centered
    finite-window spectrum of $\operatorname{Re}C_{xx}(\tau)$ (b) over $|\omega|\leq2\Omega$. Black solid: $\gamma = 0.1$; red dashed: $\gamma = 5$. Initial state: cavity $|1\rangle$, mechanical $|0\rangle$.
    Bath: OU, $\Gamma = 0.5\Omega$; $\omega_0 = 5\Omega$;
    $\lambda=0.5\Omega$ and $\kappa=0.1\Omega$.}%
    \label{fig:xx-f0}
\end{minipage}
\hfill
\begin{minipage}[t]{0.48\linewidth}
    \centering
    \captionsetup{width=\linewidth}
    \includegraphics[width=\linewidth]{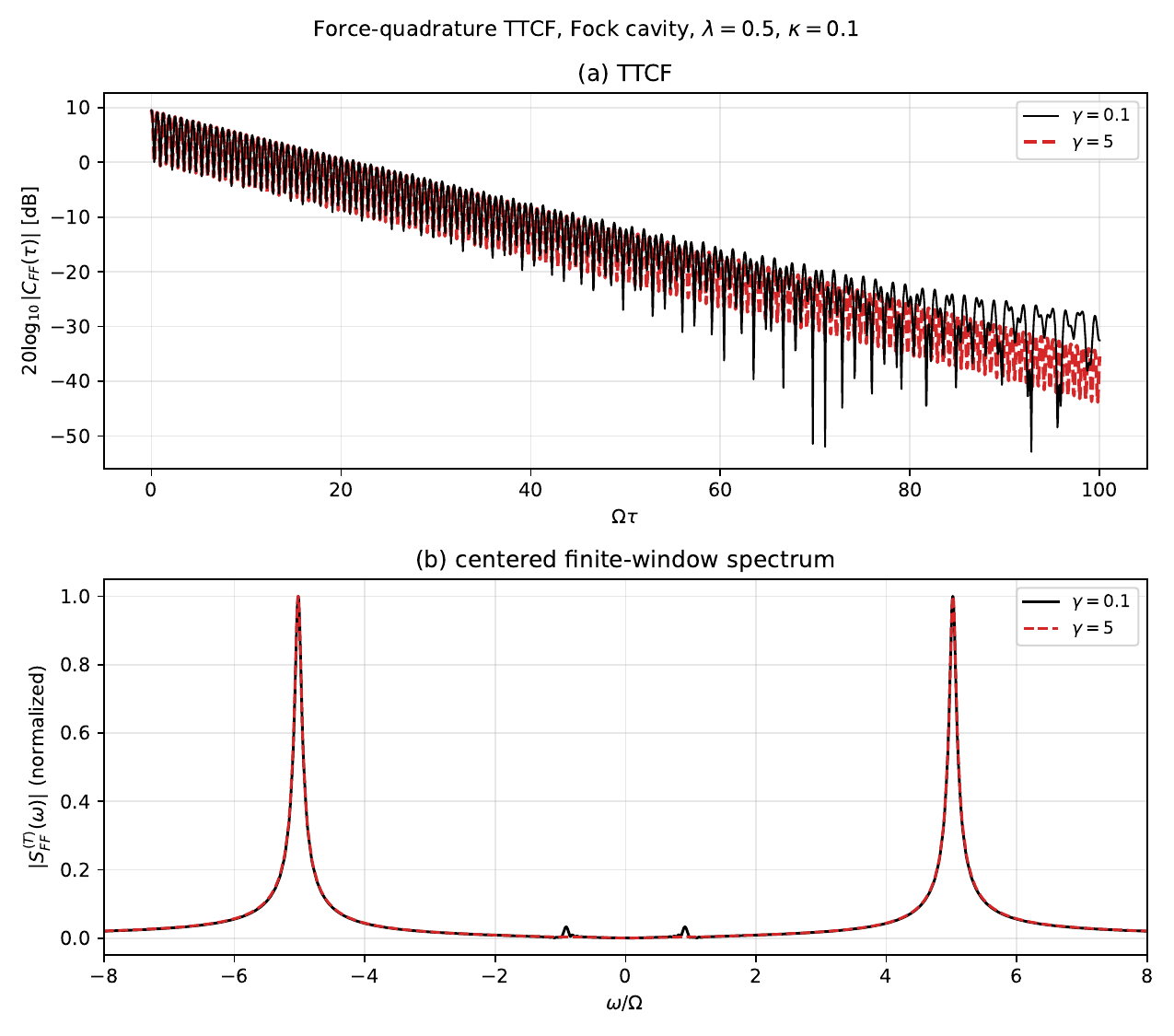}
    \captionof{figure}{%
    Force-quadrature panel: $20\log_{10}|C_{FF}(\tau)|$ (a) and the normalized, centered
    finite-window spectrum of $\operatorname{Re}C_{FF}(\tau)$ (b) over $|\omega|\leq8\Omega$.
    Black solid: $\gamma = 0.1$; red dashed: $\gamma = 5$.
    Same initial state and parameters as Fig.~\ref{fig:xx-f0}.%
    The two TTCF traces nearly overlap at early times; both curves are present in panel~(a).}%
    \label{fig:FF-f0}
\end{minipage}
\end{figure}

In the non-Markovian case ($\gamma=0.1\Omega$), $C_{xx}(\tau)$ exhibits
slowly decaying oscillations, reflecting the long correlation time of the
structured phononic environment. Its centered finite-window spectrum is
therefore dominated by a narrow pair of features near the mechanical
frequency. In the short-memory regime ($\gamma=5\Omega$), the displacement
correlation decays more rapidly and the corresponding mechanical-frequency
features become broader.

Figures~\ref{fig:xx-f0} and~\ref{fig:FF-f0} probe different
operator sectors of the same optomechanical dynamics. The displacement TTCF is
generated by the insertion $x=b+b^\dagger$ and is therefore directly sensitive
to the mechanical mode and its structured phononic environment. The
force-quadrature TTCF is generated instead by
$\hat F=a+a^\dagger$ and probes the optical sector, while the influence of the
phononic bath is transferred indirectly through the optomechanical coupling.
Although both correlations evolve under the same linear open-system map, their
operator-conditioned initial states are different. In particular, the
force-quadrature branch contains the cavity moments
$\langle a\hat F\rangle$ and
$\langle a^\dagger\hat F\rangle$. Its finite-window spectrum is consequently
dominated by optical-frequency features near
$\omega=\pm\omega_0$, which motivates the wider range
$|\omega|\leq8\Omega$ used in Fig.~\ref{fig:FF-f0}. The apparent rise previously
seen at the boundary of a mechanical-frequency window was therefore a
truncated flank of this optical structure rather than an additional resonance.
Thus, the differences between Figs.~\ref{fig:xx-f0} and~\ref{fig:FF-f0} arise
from their distinct operator insertions and dominant frequency scales, not from
different system parameters.

For the single-exponential OU kernel, the time-domain memory convolution and
the auxiliary-mode description represent the same structured-bath mechanism.
In the exact OU-pseudomode embedding, the damped auxiliary mode introduces
additional dynamical poles and finite-time interference, whereas in the QSD
description the same bath memory is carried by the memory-weighted
$O$ operator within the stated closure approximation. Observation-time and
window-function tests show that the dominant modal structure remains stable,
while fine fringes whose spacing changes with the finite record are
window-dependent. We therefore do not interpret these fringes as additional
physical resonances or independent readout channels.


\begin{figure}[htbp]
    \centering
    \includegraphics[width=0.75\linewidth]{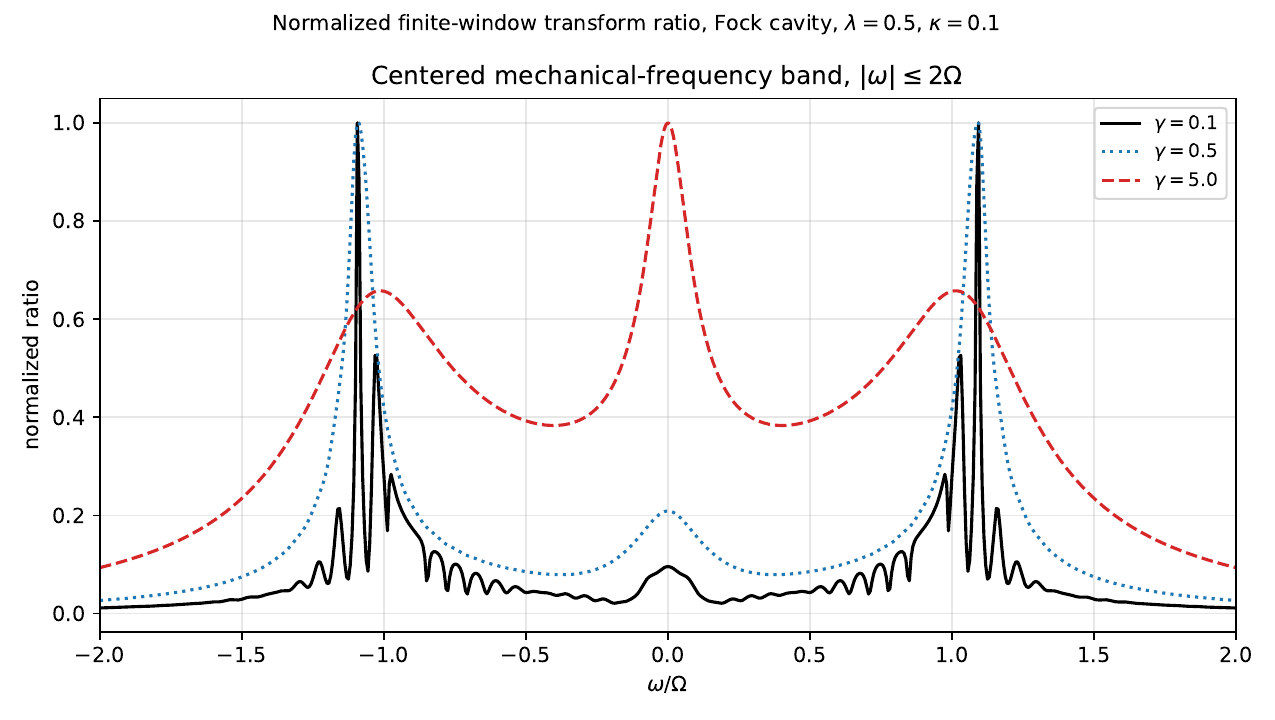}
    \caption{%
    Normalized finite-window ratio $\widehat R_T(\omega)$ for varying bath memory over the
    centered mechanical-frequency band $|\omega|\leq2\Omega$. Only frequency points
    satisfying $\widehat S_{FF}^{(T)}\geq10^{-2}$ for every displayed curve are retained.
    Black solid: $\gamma = 0.1$; blue dotted: $\gamma = 0.5$; red dashed: $\gamma = 5$.
    Initial state: cavity $|1\rangle$, mechanical $|0\rangle$.
    Bath: OU, $\Gamma = 0.5\Omega$; $\omega_0 = 5\Omega$;
    $\lambda=0.5\Omega$ and $\kappa=0.1\Omega$.}
    \label{fig:sus-f0}
\end{figure}

\textcolor{black}{Figure~\ref{fig:sus-f0} presents the normalized finite-window
ratio constructed from the independently calculated displacement and
force-quadrature spectra. Since both the numerator and the denominator vary
with the bath-memory parameter $\gamma$, the ratio is used only to compare the
relative spectral reshaping produced by different memory regimes. Because the
ratio is obtained from the centered transforms of real TTCFs, it is even in
frequency. Throughout the displayed frequency band,
$\widehat S_{FF}^{(T)}(\omega)$ remains above the prescribed denominator
threshold, so no data points are removed and no artificial floor is introduced.
The position and height of the ratio features are therefore not interpreted as
physical resonances, mechanical susceptibilities, or sensing figures of merit.
An operational interpretation of this quantity would require a separate
input--output measurement model.}

\textcolor{black}{We also repeated the calculation using a factorized coherent
cavity state while keeping the mechanical mode in the same zero-mean vacuum
state. The resulting $C_{xx}(t,0)$ and $S_{xx}^{(T)}(\omega)$ coincide with
those obtained for the factorized Fock-state preparation, so an additional
displacement figure is not shown. Changing the cavity marginal can nevertheless
modify $C_{FF}(t,0)$ and therefore the denominator of
$\widehat R_T(\omega)$. This change in the ratio originates from the
force-quadrature correlation and does not represent a dependence of the
mechanical displacement TTCF on the cavity marginal.}

As an independent validation of the two-trajectory calculation,
\textcolor{black}{we compare the QSD results for both $C_{xx}(t,0)$ and
$C_{FF}(t,0)$ with an OU-pseudomode reference. For the zero-temperature,
single-exponential OU correlation function considered here, the structured bath
can be represented exactly by one damped pseudomode}%
~\cite{Garraway1997,Tamascelli2018}:
\[
H_{\rm sys}
\rightarrow
H_{\rm sys}+g(bc^\dagger+b^\dagger c),
\qquad
g=\sqrt{\Gamma\gamma/2},
\]
with pseudomode decay rate $\Gamma_{\rm pm}=2\gamma$.
\textcolor{black}{The comparison is performed directly at the level of the two
TTCFs using the same system and bath parameters. The coupling
$\lambda=0.5\Omega$ is chosen within the parameter range in which the QSD
closure and the exact pseudomode calculation give consistent correlation
dynamics for the memory regimes considered. This benchmark tests both the
paired-trajectory construction and the accuracy of the noise-truncated
$O$-operator closure in the parameter range used in the numerical analysis.
The ratio shown in Fig.~\ref{fig:sus-f0} is not itself an exact pseudomode
identity and remains a qualitative comparison of the two independently
calculated finite-window spectra.}

\begin{figure}[htbp]
    \centering
    \includegraphics[width=0.85\linewidth]{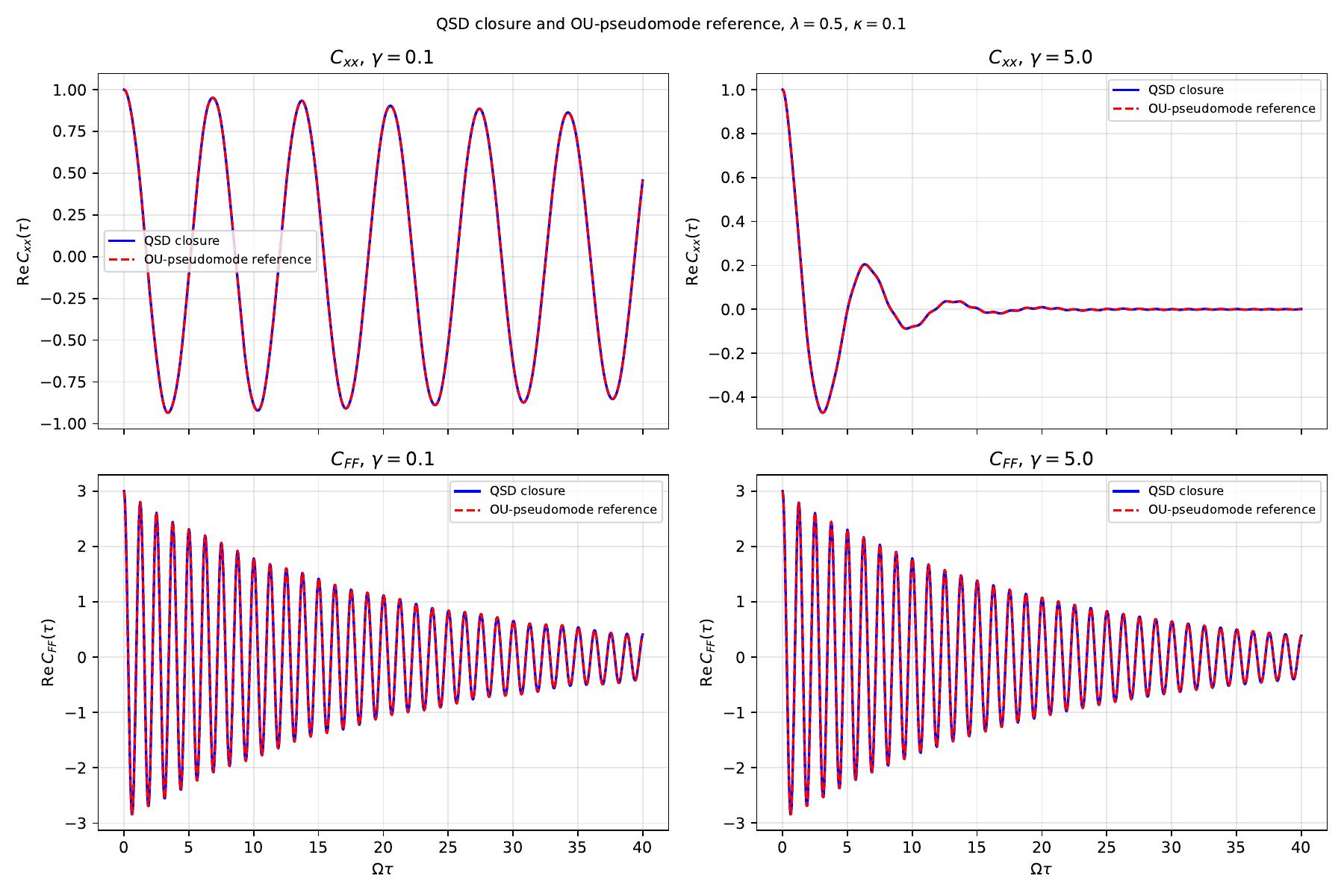}
    \caption{
Comparison of the two-trajectory QSD result (solid) with the independent
    OU-pseudomode reference (dashed; exact for the chosen OU kernel). The upper row shows
    $\mathrm{Re}\,C_{xx}(\tau)$ and the lower row shows $\mathrm{Re}\,C_{FF}(\tau)$; the left and
    right columns correspond to $\gamma=0.1\Omega$ and $\gamma=5\Omega$, respectively. Other parameters are $\kappa=0.1\Omega$,
    $\omega_0=5\Omega$, $\lambda=0.5\Omega$, and $\Gamma=0.5\Omega$.}
    \label{fig:bench}
\end{figure}

\subsection{Finite-time regression test}

\begin{figure}[htbp]
    \centering
    \includegraphics[width=0.98\linewidth]{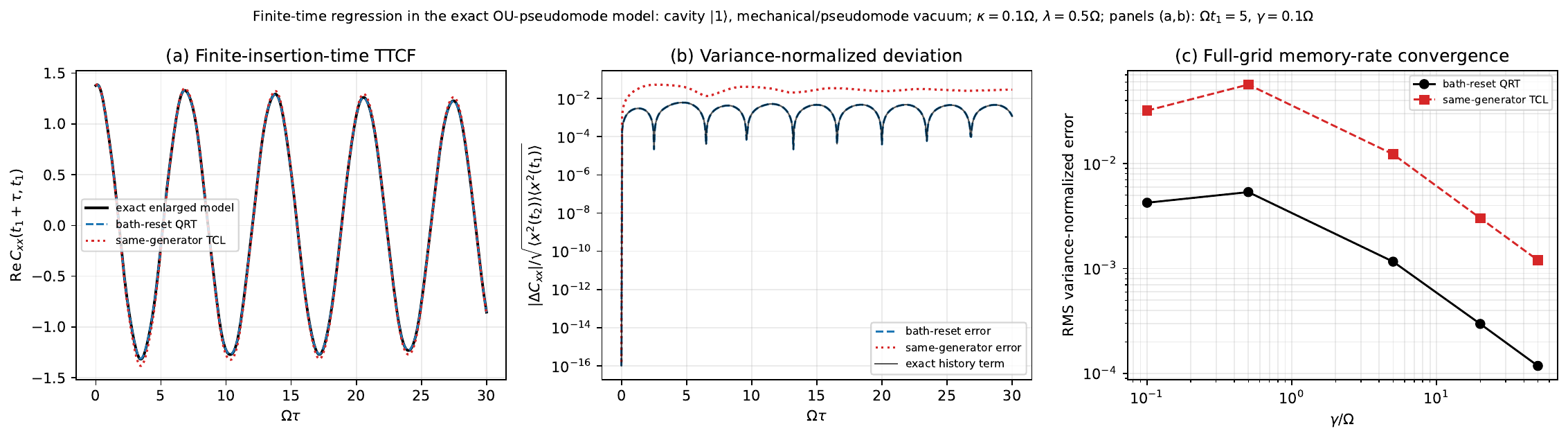}
    \caption{
Exact finite-insertion-time regression test for the OU-pseudomode model.
    (a) Real part of the exact displacement TTCF, bath-reset reduced QRT, and time-local same-generator result at
    $\Omega t_1=5$ and $\gamma=0.1\Omega$. (b) Variance-normalized deviations; the independently
    evaluated history term overlaps the bath-reset deviation. (c) Root-mean-square deviations over
    $0\leq\Omega t_1\leq20$ and $0\leq\Omega\tau\leq30$ as the bath memory rate is varied.
    The initial state is cavity $|1\rangle$ with mechanical and pseudomode vacuum. Other parameters
    are $\kappa=0.1\Omega$, $\omega_0=5\Omega$, $\lambda=0.5\Omega$, and $\Gamma=0.5\Omega$.}
    \label{fig:qrt-history}
\end{figure}

\textcolor{black}{The closure benchmark in Fig.~\ref{fig:bench} and the regression test answer different
questions. Figure~\ref{fig:qrt-history} uses only the exact OU-pseudomode model and therefore does
not contain the four-coefficient QSD approximation. Panel~(a) compares
$\mathrm{Re}\,C_{xx}(t_1+\tau,t_1)$ for the exact result, the bath-reset expression, and the
time-local same-generator expression at a representative finite
insertion time. For either reduced construction $q$, panel~(b) uses the pointwise error
\begin{equation}
\epsilon_q(t_1,\tau)=
\frac{|C_{xx}^{q}(t_1+\tau,t_1)-C_{xx}^{\rm ex}(t_1+\tau,t_1)|}
{\sqrt{\langle x^2(t_1+\tau)\rangle\langle x^2(t_1)\rangle}}.
\label{eq:variance_normalized_error}
\end{equation}
The bath-reset deviation coincides with the independently evaluated history term after the same
normalization.}

\textcolor{black}{The representative curve is accompanied by a scan over the full displayed $t_1$--$\tau$
grid. Panel~(c) reports $[N^{-1}\sum_{t_1,\tau}\epsilon_q^2]^{1/2}$ over that grid rather than
selecting the largest separation at one insertion time. The deviations vanish in the Markov limit
and decrease for sufficiently large $\gamma$; no monotonic dependence is inferred from the
intermediate memory rates. At $t_1=0$ they coincide identically with the exact embedded correlation
because the factorized preparation and zero-mean pseudomode vacuum give a vanishing initial
system--pseudomode cross moment. The finite-$t_1$
separation therefore tests the missing environmental history, not the absence of a steady-state
Fourier transform.}

\subsection{Initially correlated cavity-mechanical state}
\label{sec:entangled_initial}

\textcolor{black}{The factorized-state results show that changing the cavity
marginal alone does not modify the displacement TTCF when the mechanical mode
is prepared in the same zero-mean vacuum state. We therefore next examine
whether joint cavity--mechanical correlations can alter $C_{xx}(t,0)$ while
the local cavity and mechanical marginal states are kept fixed.}

\begin{figure}[htbp]
    \centering
    \includegraphics[width=0.98\linewidth]{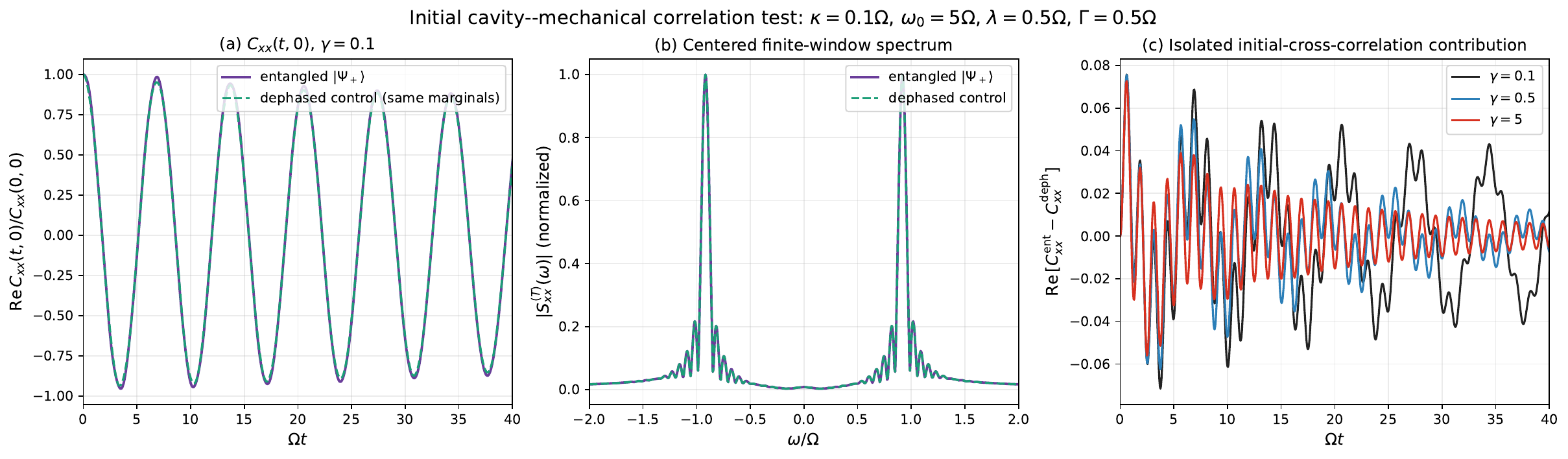}
    \caption{
Initial cavity--mechanical correlation test. (a)
    $\mathrm{Re}\,C_{xx}(t,0)/C_{xx}(0,0)$ for $|\Psi_+\rangle$ and the dephased same-marginal
    control at $\gamma=0.1\Omega$, shown over $0\leq\Omega t\leq40$. (b) Normalized centered
    finite-window spectra of $\operatorname{Re}C_{xx}$ calculated with $T=100/\Omega$.
    (c) Isolated real initial-cross-correlation contribution
    $\mathrm{Re}[C_{xx}^{\rm ent}(t,0)-C_{xx}^{\rm deph}(t,0)]$ for three memory regimes.
    Parameters are $\kappa=0.1\Omega$,
    $\omega_0=5\Omega$, $\lambda=0.5\Omega$, and $\Gamma=0.5\Omega$.}
    \label{fig:Cxx-entangled}
\end{figure}

To isolate the effect of joint initial correlations, we compare the entangled state
\begin{equation}
|\Psi_+\rangle=
\frac{|1_a0_b\rangle+|0_a1_b\rangle}{\sqrt{2}}
\end{equation}
with the dephased state
\begin{equation}
\rho_{\rm deph}
=
\tfrac12|1_a0_b\rangle\langle1_a0_b|
+
\tfrac12|0_a1_b\rangle\langle0_a1_b|.
\end{equation}
The two states have identical cavity and mechanical marginals and the same
$C_{xx}(0,0)=2$, but differ in their initial cross moments:
$\langle ax\rangle_0=\langle a^\dagger x\rangle_0=1/2$ for
$|\Psi_+\rangle$, while both vanish for $\rho_{\rm deph}$. The separation in
Fig.~\ref{fig:Cxx-entangled} therefore arises from the joint initial
correlations. Panel~(c) isolates
$\mathrm{Re}[C_{xx}^{\rm ent}(t,0)-C_{xx}^{\rm deph}(t,0)]$, while panel~(b)
shows the corresponding centered finite-window spectra.

The difference is stable under numerical refinement, satisfies the analytic
cross-moment relation, and is reproduced by the OU-pseudomode calculation.
Thus, changing only the cavity marginal does not modify $C_{xx}$ for
factorized states with the same zero-mean mechanical vacuum, whereas relevant
cavity--mechanical cross correlations can restore initial-state dependence.
This conclusion is not specific to entanglement, since classically correlated
states may contain the same cross moments.

\begin{figure}[htbp]
    \centering
    \includegraphics[width=0.98\linewidth]{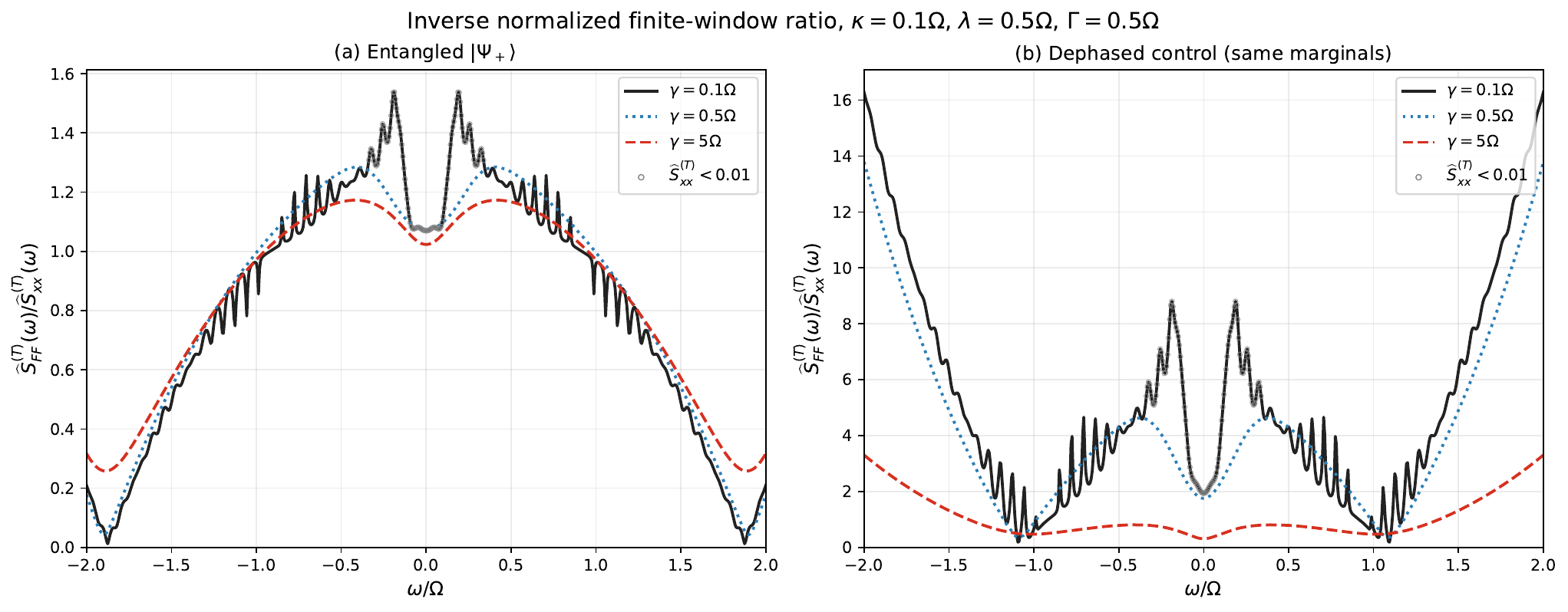}
    \caption{
Inverse normalized finite-window ratio
    $\mathcal I_T(\omega)=\widehat S_{FF}^{(T)}(\omega)/\widehat S_{xx}^{(T)}(\omega)$ for
    (a) the entangled state $|\Psi_+\rangle$ and (b) its dephased same-marginal control. Black solid,
    blue dotted, and red dashed curves correspond to $\gamma/\Omega=0.1$, $0.5$, and $5$,
    respectively. Every sampled frequency in $|\omega|\leq2\Omega$ is retained; no denominator
    threshold, clipping, or post-division normalization is applied. Open gray markers identify
    $\widehat S_{xx}^{(T)}<10^{-2}$. The observation window is $T=100/\Omega$. Other parameters are $\kappa=0.1\Omega$,
    $\omega_0=5\Omega$, $\lambda=0.5\Omega$, and $\Gamma=0.5\Omega$.}
    \label{fig:entangled-ratio}
\end{figure}

\textcolor{black}{Figure~\ref{fig:entangled-ratio} shows the inverse
finite-window ratio
\[
\mathcal I_T(\omega)
=
\frac{\widehat S_{FF}^{(T)}(\omega)}
     {\widehat S_{xx}^{(T)}(\omega)}
\]
for the entangled state and its dephased same-marginal control. The two
spectra are normalized independently over $|\omega|\leq2\Omega$ before the
quotient is formed, and $\mathcal I_T(\omega)$ is not subsequently rescaled.
All sampled frequencies are displayed. Open gray markers indicate the region
where $\widehat S_{xx}^{(T)}(\omega)<10^{-2}$; large or sharp ratio features
at these points are therefore attributed to division by a small displacement
spectrum rather than to additional dynamical resonances. Away from the marked
region, the difference between the two profiles reflects the influence of the
initial cavity--mechanical cross correlations on the underlying TTCFs and
their finite-window transforms. Accordingly, $\mathcal I_T(\omega)$ is used
only as a qualitative comparison of relative spectral shapes, not as a
susceptibility, entanglement witness, or sensing figure of merit.}

\section{Conclusion}
\label{sec:conclusion}

Building on our stochastic-trajectory formulation of non-Markovian
two-time correlation functions, we have extended the method
to a continuous-variable cavity optomechanical system coupled to two
environments with distinct memory properties. The physical and
operator-conditioned branches are propagated with the same phononic-noise
realization, so that the environmental history associated with the first
operator insertion is retained in the TTCF calculation. The present extension
does not assume stationarity or invoke the standard same-generator form of the
quantum regression theorem. For the Ornstein--Uhlenbeck mechanical bath, the
memory convolution is reduced to a model-specific closed set of time-local
auxiliary equations, while the construction can be generalized to structured
bath correlations represented by sums of exponentials.

Applied to a linearized cavity optomechanical system with a structured
phononic bath and Markovian optical decay, the framework reveals the
qualitative signatures of finite environmental memory in both temporal and
spectral correlations. Long bath memory produces persistent oscillatory TTCFs
and structured, non-Lorentzian finite-window spectra, whereas the
short-memory regime exhibits more rapidly decaying correlations and broader
spectral features. The displacement and force-quadrature TTCFs probe different
operator sectors and therefore display different dominant frequency
structures under the same open-system dynamics. The independent
OU-pseudomode benchmark confirms the validity of the noise-truncated
$O$-operator closure in the parameter regime considered, while the
finite-time regression test explicitly identifies the contribution
of environmental history that is absent from reduced same-generator
constructions.

The analysis also distinguishes environmental-memory effects from
initial-state effects. For factorized preparations with the same zero-mean
mechanical vacuum, changing only the cavity marginal leaves the displacement
TTCF unchanged. Initial cavity--mechanical cross correlations, however, enter
the displacement correlation directly and can modify its finite-window
spectral response even when the local marginal states are identical. Taken
together, these results establish a history-preserving and internally
validated framework for studying dynamical correlations in structured
environments. The independently calculated displacement and
force-quadrature spectra, and their regularized ratios, are used here as
theoretical diagnostics of the underlying correlation dynamics. Future work
may extend the approach to finite-temperature and more general
non-Lorentzian reservoirs, improved non-Markovian closure schemes, and a full
input--output measurement model connecting the calculated TTCFs to
experimentally detected signals.

\begin{backmatter}
\bmsection{Funding} National Science Foundation (2328948, 2519355, 2519356).

\bmsection{Disclosures} The authors declare no conflicts of interest.

\bmsection{Data availability} Data underlying the results presented in this paper are available from the corresponding author upon reasonable request.
\end{backmatter}

\bibliographystyle{opticajnl}
\bibliography{reference}

\end{document}